\documentclass{article}
\usepackage{graphicx}
\def\DR{\rm I\kern-1.45pt\rm R}
\def\DC{\kern2pt {\hbox{\sqi I}}\kern-4.2pt\rm C}

\newcommand{\ba}{\begin{array}}
\newcommand{\ea}{\end{array}}
\newcommand{\be}{\begin{equation}}
\newcommand{\ee}{\end{equation}}
\newcommand{\bea}{\begin{eqnarray}}
\newcommand{\eea}{\end{eqnarray}}
\newcommand{\bi}{\begin{itemize}}
\newcommand{\ei}{\end{itemize}}

\usepackage{amscd,amsmath,amssymb}

\begin{document}\begin{center}
{\bf \Large Exactly solvable Ising--Heisenberg chain with triangular $XXZ$-Heisenberg plaquettes.  }\\
\vspace{0.5 cm} {\large Diana Antonosyan$^{1,2}$, Stefano
Bellucci$^3$ and Vadim Ohanyan$^{1,4}$}
\end{center}
\noindent $\;^1${\it Department of Theoretical Physics, Yerevan State University, A.Manoogian, 1,
Yerevan,
0025 Armenia}\\
$\;^2$ {\it Institute for Physical Research of National Academy of Sciences of Armenia, Ashtarak-2, 0203, Ashtarak, Armenia}\\
$\;^3$ {\it INFN-Laboratori Nazionali di Frascati, Via E. Fermi 40,
00044 Frascati, Italy}\\
 $\;^4${\it Yerevan Physics Institute, Alikhanian Br.2,
Yerevan, 0036, Armenia}\\
{\sl E-mails: bellucci@lnf.infn.it,
  ohanyan@yerphi.am}
\begin{abstract}
 A mixed Ising-Heisenberg spin system
consisting of triangular $XXZ$-Heisenberg spin clusters assembled
into a chain by alternating with Ising spins interacting to all
three spins in the triangle is considered. The exact solution of the
model is given in terms of the generalized decoration--iteration map
and within the transfer-matrix technique. Exact expressions for
thermodynamic functions are derived. Ground state phase diagrams,
thermodynamic and magnetic properties of the system are examined.
\end{abstract}

\section{Introduction}

    Frustrated spin systems have been in the focus of intensive
 investigations during the last decades \cite{frus}. Belonging to
 a separate class of magnetic systems, they possess a number of
 fascinating features. Geometrically frustrated magnets are of
 special interest due to their unusual magnetic and thermodynamic
 properties. In this class of magnets the lattice topology precludes
 the spins from simultaneously minimizing all spin--spin--exchange
 interactions. The simplest example of frustrated spin cluster  is given by the
 triangular arrangement of three spins with antiferromagnetic
 interaction between each pair.

    Numerous variants of lattice spin
 systems containing triangular plaquettes are, on one hand,
 widespread structures of magnetic materials and provide, on the other hand,
 a theoretical prototype models for investigating the
 geometrically frustrated systems and their unusual features. Among
 the important classes of geometrically frustrated two-dimensional systems, it is
 worth mentioning the triangular antiferromagnets \cite{col97} and
 antiferromagnets on kagom$\acute{e}$ lattice \cite{kag}. There are
 few one-dimensional lattices with triangular frustrated units which
 are famous for their exactly known dimerized ground states: the
 Majumdar-Ghosh model which is the special case of spin-$1/2$ chain
 with competing nearest-neighbor (NN) and next--nearest--neighbor (NNN) interaction \cite{MG} and the so--called
 sawtooth chain ($\Delta$-chain), the system of corner--sharing triangles \cite{saw}.

    Although, properties of the ground state of the
    above--mentioned systems have been investigated analytically very
    well, only numerical calculations still remains reliable to gain an
    insight into the finite $T$ thermodynamic properties. On the
    other hand, there is a large class of exactly solvable one
    dimensional "classical" lattice models which allow one to obtain an
    exact expressions for thermodynamic functions using different
    techniques \cite{Bax}. The simplest and at the same time key example is
    the transfer-matrix solution of the one-dimensional Ising model
    known since 1920s. Further developments of the transfer-matrix
    technique which is very powerful and rather simple for the
    arbitrary one dimensional lattice systems with discrete
    commuting variables turned out to be especially useful in the
    statistical mechanics of macromolecules\cite{macromol}. However, in the theory
    of magnetism and strongly correlated systems one-dimensional
    systems with classical variables have rather "pedagogical" value
    than practical one.

     Nevertheless, it was established in a series of recent publications
     that considering the Ising counterparts of some
     Heisenberg one--dimensional spin systems one can obtain, on
     one hand, the exact thermodynamic description of the
     problem and, on the other hand, the system which at least on the
     qualitative level exhibits the magnetic and thermodynamic
     behaviors very similar to those of the underlying quantum spin
     model \cite{ffa}-\cite{lad}. For instance, the form of the
     magnetization curve does not acquire crucial changes when one
     replaces some or even all Heisenberg interactions in the chain
     with Ising ones. So, if the initial quantum spin chain exhibits a
     magnetization plateau at a certain value of the magnetization
     this feature will hold for the corresponding Ising or
     Ising--Heisenberg chain, whereas quantitative characteristics of
     the plateau (terminal points, width, e.t.c.) can be
     different. This feature was reported in Ref. \cite{ffa}
     for the spin-1/2 chain with bond alternating
     ferromagnetic--ferromagnetic--antiferromagnetic (F-F-AF) interaction,
     the model of 3CuCl$_2\cdot$ 2dx (dx=1,4-dioxane) compound. In the quantum
     model describing the magnetic structure of 3CuCl$_2\cdot$ 2dx,
     considered numerically in Ref. \cite{hida} the
     appearance of magnetization plateau at $1/3$ of the saturation magnetization value was
     established. It is worthy to note that experiments have
      shown no plateau in 3CuCl$_2\cdot$ 2dx which is caused by the insufficient small ratio of
      the antiferromagnetic and ferromagnetic coupling constants.
       The corresponding regime of magnetic behavior for F-F-AF Ising chain was also established in
       Ref. \cite{ffa} This was the example of the intermediate
     magnetization plateaus in one--dimensional spin system which in
     recent decades received considerable attention both in
     theoretical and experimental aspects\cite{frus,pla}. Hereafter, the
     Ising counterpart of this model was solved exactly in thermodynamic context in Ref.
     \cite{ffa} demonstrating the same magnetization plateau at
     $m=1/3$. Almost the same program but with different
     techniques was performed in Ref. \cite{str2}
     for another bond alternating chain, the spin-1/2
     F-F-AF-AF chain which is regarded as the model of magnetic structure of
     Cu(3-Clpy)$_2$(N$_3$)$_2$, where 3-Clpy indicates
     3-chloropyridine.
     The authors considered a simplified Ising--Heisenberg
     variant of the F-F-AF-AF chain within the method of exact mapping
     transformation (decoration--iteration transformation)\cite{mt}.
     Comparing their results with the experimental and numerical
     data they found not only qualitative but also quantitative
     agreement as well. It is also worthy to mention the works on
     exact solutions of mixed spin-(1/2,1,3/2) and spin-1/2, the so called diamond
     chains with the decoration--iteration transformation technique
     \cite{str}, which revealed series of magnetization plateaus in
     magnetization process of the system. Also, the magnetization
     plateaus in one-dimensional spin-1, spin-3/2 and spin-2 Ising chains with single--ion anisotropy
      have been investigated in Ref. \cite{ayd} showing a good
      agreement with experimental data obtained for spin-1
      Ni-compounds. Magnetization plateaus in the Ising limit of the
      multiple--spin--exchange model on the four-spin cyclic
      interaction have been considered in Ref. \cite{Iran}.

      All these results, unexpected at the first glance, offer the
     challenge to develop a formalism which, on one hand, leads
     to the exact analytical treatment for thermodynamic functions of
     the one--dimensional spin systems and, on the other hand, could
     be relevant in describing and understanding experimental
     data  for real materials. As mentioned above, the idea is
     very simple, just to replace some (or even all) Heisenberg-type
     quantum interaction with the Ising ones. Of course, there are
     many very important features of quantum exchange interaction
     which are crucial in understanding many phenomena in
     magnetic systems and which cannot be neglected. But the
     results of Ref. \cite{ffa}-\cite{Iran} indicate the
     existence of at least a few quantum spin chains (actually
     the chains with alternating ferromagnetic and antiferromagnetic
     couplings), the thermodynamic properties of which do not undergo
     sizable changes under the replacement of some interaction with
     more simple Ising ones, allowing an exact solution. In addition to that, the spin systems considered here may serve
     as more or less realistic model for the arrays of single--molecule magnets (SMMs) \cite{smm} with inter--molecular couplings, engineered
     by the methods of the so--called supra--molecular chemistry \cite{smc}. SMMs recently receive much attention as they are very convenient objects for studying the phenomena which are critical at the nanoscale, such as quantum tunneling of magnetization, decoherence, and Berry phases. One of the well--studied specific examples of SMMs, the equilateral spin triangle Cu$_3$ \cite{cu3}, assembled to the one-dimensional arrays with very weak inter-molecular coupling could be the system which approximately can be described by the model, considered here.

      In this paper we consider a further development of the methods
      applied in the Ref. \cite{ffa}-\cite{lad}, for the chain
      with frustrated triangular quantum XXZ-Heisenberg plaquettes.
      This chain was introduced in Ref. \cite{roj} where the ground
      state properties and some numerical results for purely
      Heisenberg case were obtained.
      Within the transfer-matrix technique we obtain exact thermodynamic solution of the chain
      consisting of such spin triangles alternating with the single
      Ising spins which are coupled by the interaction of Ising type
      with all six spins of the sites of two neighbor triangular
      plaquettes. The paper is organized as follows. In the Second
      section we formulate the model and present its solution by the
      decoration--iteration transformation technique. In the Third
      section the transfer-matrix solution of the system is given.
      In the Next section, using the transfer-matrix formalism, the exact
      thermodynamics of the system is discussed. In
      particular the $T=0$ phase diagram and plots of magnetization
      vs. external magnetic field, specific heat and magnetic
      susceptibility are presented. The Appendix contains several
      technical points of the calculations.

 \section{The model and its solution with the aid of the generalized decoration--iteration transformation.}
\begin{figure}
 \begin{center}
  \includegraphics[scale=0.7]{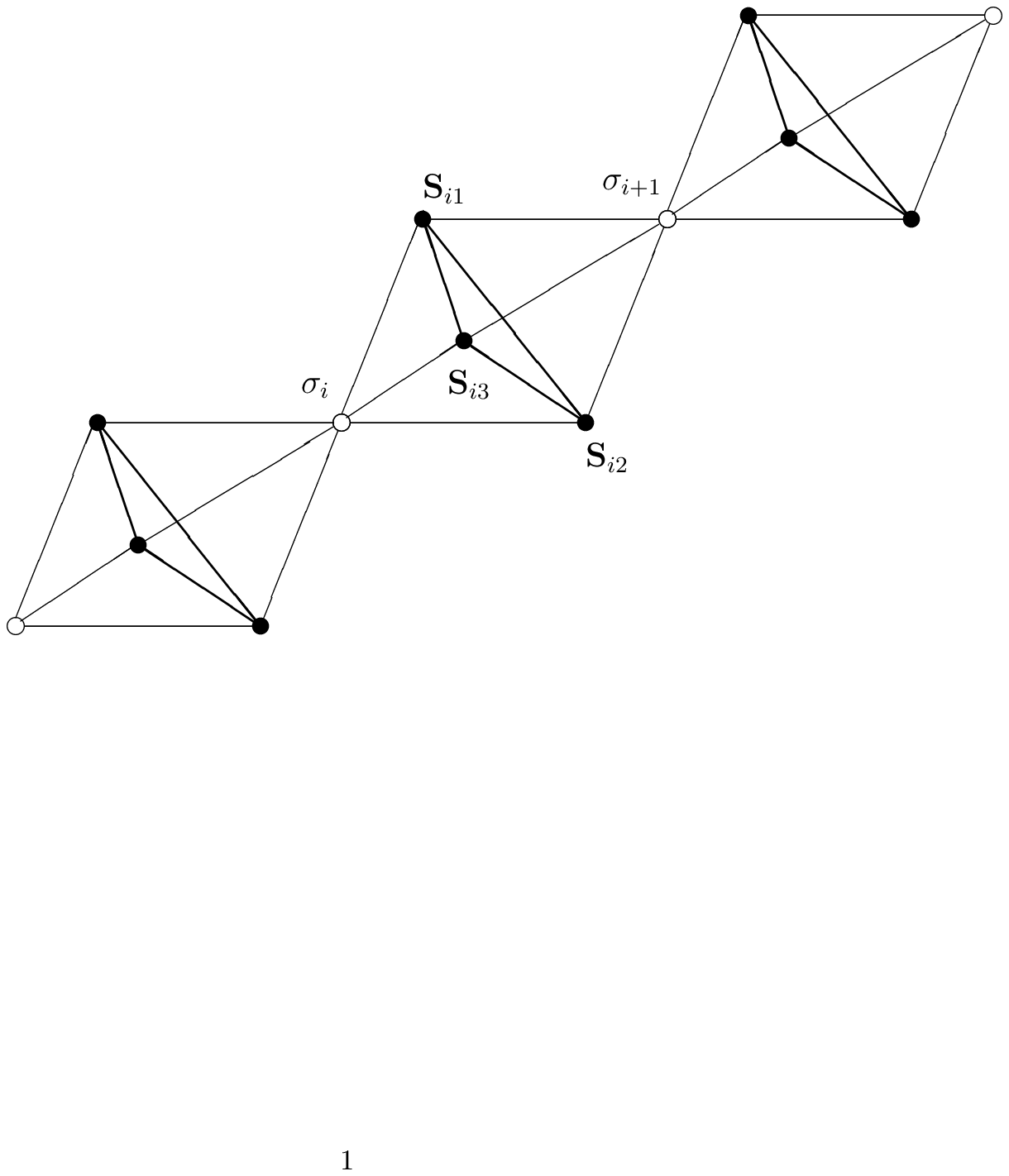}
  \caption{The schematic picture of the Ising--Heisenberg chain with triangular plaquettes alternating with single sites.
  The filled(empty) circles indicate Heisenberg(Ising) spins.\label{fig1}}
           \end{center}
         \end{figure}

 We begin with the hybrid Ising-Heisenberg spin system consisting of
 the triangular clusters of Heisenberg $S=1/2$ spins interacting to
 each other with coupling constant $J$ and axial anisotropy
 $\Delta$. These triangles are assembled into a chain by
 alternating with single sites. Each spin at the single site is coupled to all
 six spins situated at the corners of its two adjacent triangles (see
 Fig.\ref{fig1}). The interactions of spins at single sites and spins
 belonging to triangles are all supposed to be of Ising type (the interaction
 includes only $z$-components of the spins). The
 corresponding Hamiltonian is suitable to be presented as a sum over the
 plaquette Hamiltonians each one containing one triangle and its two
 surrounding single sites:
 \bea
 &&{\mathcal{H}}=\sum_{i=1}^N{\mathcal{H}}_i, \label{H} \\ \nonumber
&&{\mathcal{H}}_i=J[\frac{\Delta}{2}(S_{i1}^+ S_{i2}^-+S_{i1}^-
S_{i2}^+)+S_{i1}^zS_{i2}^z+\frac{\Delta}{2}(S_{i1}^+
S_{i3}^-+S_{i1}^-
S_{i3}^+)+S_{i1}^zS_{i3}^z+\frac{\Delta}{2}(S_{i2}^+
S_{i3}^-+S_{i2}^- S_{i3}^+)+S_{i2}^zS_{i3}^z]\\ \nonumber
&&+K(S_{i1}^z+S_{i2}^z+S_{i3}^z)(\sigma_i+\sigma_{i+1})-H_2(S_{i1}^z+S_{i2}^z+S_{i3}^z)-\frac{H_1}{2}(\sigma_i+\sigma_{i+1}).
 \eea
Here by $\sigma$ we denote the $z$-components of the spins at the
single sites which, thus, can be identified with the Ising variables
taking values $\pm 1$, $K$ is Ising coupling between triangle spins
and single site spins, $H_1$ and $H_2$ stand for the couplings of
 Ising and Heisenberg spins to the external magnetic field pointing
in $z$ direction. For convenience we normalize the Heisenberg spin
operators in such a way which provides the eigenvalues of $S^z$ to
take $\pm 1$ values, which means that our $S^{\alpha}$ are just the Pauli matrices without spin magnitude multiplier. (In order to recover conventional Heisenberg Hamiltonian parameters one should replace in all our formulas $J$ with $J/4$ and $H$ with $H/2$)  Thus, they obey the commutation
relations: \bea &&[S_{i a}^+, S_{j
b}^-]=\delta_{i j} \delta_{a b} 4 S_{i a} ^z, \label{com} \\
\nonumber &&[S_{i a}^z, S_{j b}^{\pm}]=\pm 2 \delta_{i j} \delta_{a
b} S_{i a} ^{\pm}, \eea and act on the basic states in the following
way \bea &&S^z|\uparrow \rangle=|\uparrow \rangle, \quad
S^z|\downarrow \rangle=-|\downarrow \rangle, \label{act} \\
&&\nonumber S^+|\uparrow \rangle=0, \quad S^+|\downarrow
\rangle=2|\uparrow \rangle \\ \nonumber && S^-|\uparrow
\rangle=2|\downarrow \rangle, \quad S^-|\downarrow \rangle=0.
 \eea
 In order to describe thermodynamics of the system one need to
 calculate the partition function
 \bea
 {\mathcal{Z}}=\sum_{(\sigma)}\mbox{Tr}_{(S)}e^{-\beta\sum_{i=1}^N{\mathcal{H}}_i},
 \label{z1}
 \eea
 where the sum is going over all possible configurations of the
 Ising spins and $\mbox{Tr}_{(S)}$ denotes the trace over all
 Heisenberg operators $S$ and $\beta$ as usually is the inverse temperature. One can easily see that the
 Hamiltonians
 corresponding to different plaquettes do commute which allows one
 to expand the exponent in the partition function obtaining partial factorization of the partition function
 \bea
{\mathcal{Z}}=\sum_{(\sigma)}\prod_{i=1}^N \mbox{Tr}_i
e^{-\beta{\mathcal{H}}_i}, \label{z2}
 \eea
 here $\mbox{Tr}_i$ stands for the trace over the state of ith
 triangle. Now we can utilize the so--called decoration-iteration
 transformation \cite{str} which allows us to represent the trace over
 all states of single triangle interacting with two Ising spin as
 the elements of the transfer-matrix for ordinary one-dimensional
 Ising model with a certain renormalization of its parameters:
 \bea
\mbox{Tr}_i e^{-\beta {\mathcal{H}}_i}=\sum_{n=1}^8 e^{-\beta
\lambda_n(\sigma_i, \sigma_{i+1})}= Ae^{\beta R \sigma_i
\sigma_{i+1}+\beta\frac{H}{2}(\sigma_i+\sigma_{i+1})}, \label{DIT}
 \eea
 here $\lambda_n(\sigma_i, \sigma_{i+1})$ are eight eigenvalues of
 the ${\mathcal{H}}_i$ which depend of the values of its neighbor
 Ising spins. The parameters of emergent "one-dimensional Ising"
 model are expressed by the parameters of the initial model in the
 following way:
 \bea
 A=\left(Z_+Z_-Z_0^2 \right)^{1/4}, \quad \beta
 R=\frac{1}{4}\log\left( \frac{Z_+Z_-}{Z_0^2}\right), \quad \beta
 H=\beta H_1+\frac{1}{2}\log\left( \frac{Z_+}{Z_-} \right),
 \label{par}
 \eea
 where
 \bea
 Z_+=\sum_{n=1}^8 e^{-\beta
\lambda_n(+, +)}, \quad  Z_-=\sum_{n=1}^8 e^{-\beta \lambda_n(-,
-)}, \quad  Z_0=\sum_{n=1}^8 e^{-\beta \lambda_n(+, -)}=\sum_{n=1}^8
e^{-\beta \lambda_n(-, +)} \label{zzz}
 \eea
 are the partition functions for single triangle calculated for
 certain values of the its neighbor Ising spins. The explicit form of
 these functions can be found in the Appendix. Thus, according to
 Eqs.(\ref{z1})-(\ref{zzz}) one can obtain an exact solution for the
 system under consideration in the form corresponding to the well
 known solution of one dimensional Ising model as follows:
\bea {\mathcal{Z}}(\beta; J, K, \Delta, H_1,
H_2)=\sum_{(\sigma)}\prod_{i=1}^N Ae^{\beta R \sigma_i
\sigma_{i+1}+\beta\frac{H}{2}(\sigma_i+\sigma_{i+1})}=A^N
{\mathcal{Z}}_0(\beta; R, H), \label{z_is}
 \eea
 The partition function of the one-dimensional Ising model with periodic boundary conditions in
 thermodynamic limit is \cite{Bax}
 \bea
{\mathcal{Z}}_0(\beta; R, H)=e^{\beta N R}\left(\cosh(\beta
H)+\sqrt{\sinh^2(\beta H)+e^{-\beta 4 R}} \right)^N. \label{z0}
 \eea
 The relation (\ref{z_is}) establishes a connection between
 thermodynamic quantities of the system under consideration and
 those of one dimensional Ising model. So, in the thermodynamic limit for
 free energy per spin we get
 \bea
 f=-T \lim_{N \to \infty} (\frac{\log {\mathcal{Z}}}{4N})=\frac{1}{4}(f_0-T\log
 A)
 \label{fe}
 \eea
 (We remind that by $N$ we denote the number of the lattice
 blocks each one containing five spins, so under the periodic boundary
 conditions the total number of spins is $4N$.) Here by $f_0$ we
 denote the corresponding free energy per spin for ordinary Ising chain.
 Now we can utilize the common thermodynamic relations to obtain
 expressions for various thermodynamic quantities we are going to
 investigate. As usual for specific heat, entropy, total magnetization and susceptibility per spin one has
 \bea
C_H=-T\left(\frac{\partial^2 f}{\partial T ^2} \right)_H, \quad
S=-\left(\frac{\partial f}{\partial T} \right)_H, \quad
M=-\left(\frac{\partial f}{\partial H} \right)_T, \quad
\chi=\left(\frac{\partial^2 f}{\partial H ^2} \right)_T \label{CS}
 \eea
 The thermal averages of various microscopic variables can also be
 expressed in term of corresponding quantities for the Ising model. The
 sublattice magnetization and various nearest neighbor correlation
 functions can be expressed by the sublattice magnetization and nearest neighbor correlation functions of
 ordinary Ising chain given by
 \bea
 &&M_0=-\left(\frac{\partial f_0}{\partial H} \right)_T=\frac{\sinh(\beta H)}{\sqrt{\sinh^2(\beta H)+e^{-\beta 4 R}}},
 \label{av_i} \\ \nonumber
 &&C_0=\langle \sigma_i \sigma_{i+1}\rangle=-\left(\frac{\partial f_0}{\partial R} \right)_{T,H}
 =1-\frac{2 e^{-\beta 4 R}}{\left(\cosh(\beta
H)+\sqrt{\sinh^2(\beta H)+e^{-\beta 4 R}}\right)\sqrt{\sinh^2(\beta H)+e^{-\beta 4 R}}} \\
 \eea
For sublattice and total magnetizations for the model under
consideration one obtains:
 \bea
 &&M_1=\frac{\langle\sum_{i=1}^N \sigma_i\rangle}{N}=-4\left(\frac{\partial f}{\partial H_1}
 \right)_T=-\left(\frac{\partial f_0}{\partial H}
 \right)_T=M_0, \label{av} \\ \nonumber
  &&M_2=\frac{\langle\sum_{i=1}^N
\sum_{a=1}^3S_{ia}^z\rangle}{3N}=-\frac{4}{3} \left( \frac{\partial
f}{\partial
 H_2}\right)_T=\frac{1}{3}\left(\frac{T}{A}\left(\frac{\partial A}{\partial H_2} \right)_T
 -\left(\frac{\partial f_0}{\partial R} \right)_T\left(\frac{\partial R}{\partial H_2} \right)_T-
 \left(\frac{\partial f_0}{\partial H} \right)_T\left(\frac{\partial H}{\partial H_2}
 \right)_T\right)= \\ \nonumber
 &&\frac{1}{12}\left(W_++2W_0 M_0+W_- C_0
 \right), \\ \nonumber
 &&M=\frac{1}{4}M_1+\frac{3}{4}M_2,
 \eea
 where the corresponding functions $W_{\alpha}$, $\alpha= \pm, 0$ can be found
 in Appendix.
 In the same manner for nearest neighbor correlation functions we
 obtain
 \bea
 &&C_{\sigma \sigma}=\langle \sigma_i \sigma_{i+1}\rangle=C_0, \label{cor} \\
 \nonumber
&& C_{S \sigma}=\langle S_{ia}^z
\sigma_i\rangle=\frac{\langle\sum_{i=1}^N(\sigma_i+\sigma_{i+i})(S_{i1}^z+S_{i2}^z+S_{i3}^z)\rangle}{6N}=\frac{2}{3}\left(\frac{\partial
f}{\partial K} \right)_{T,H}=-\frac{1}{12}\left(V_++2V_-M_0+V_+C_0
\right), \\ \nonumber &&C_{SS}^{(x,y)}=\langle
S_{ia}^xS_{ib}^x\rangle=\langle
S_{ia}^yS_{ib}^y\rangle=\frac{2}{3J}\left(\frac{\partial f}{\partial
\Delta} \right)_{T,H}=-\frac{1}{12 J}\left(U_++2 U_0 M_0+U_- C_0
\right), \quad a \neq b, \\ \nonumber &&C_{SS}^z=\langle S_{ia}^z
S_{ib}^z\rangle=\frac{4}{3} \left(\frac{\partial f}{\partial J}
\right)_{T,H}-2 \Delta
C_{SS}^{(x,y)}=\frac{1}{12}\left(\frac{2\Delta}{J} U_+-F_++2(\frac{
2\Delta}{J} U_0-F_0)M_0+(\frac{2\Delta}{J} U_--F_-)C_0 \right),\quad
a \neq b.
 \eea
 \section{Solution within the transfer-matrix technique}
 We will use another technique for describing the thermodynamics of
 the system under consideration. Namely, having the analytical
 expression for the single quantum triangle (see the Appendix), one can represent the partition function of the chain
 under consideration
 given by Eq. (\ref{z2}), in the form which mimics the partition function of
 the classical chain with two state variables at each site (the "Ising chain" with arbitrary Boltzmann
 weights):
 \bea
{\mathcal{Z}}=\sum_{\sigma} \prod_{i=1}^N e^{\frac{\beta}{2}H_1
\left(\sigma_i+\sigma_{i+1} \right)}Z \left(\sigma_i,\sigma_{i+1}
\right)=\mbox{Tr}{\mathbf{T}}^N, \label{Z_T}
 \eea
where $Z \left(\sigma_i,\sigma_{i+1} \right)$ is calculated in the
Appendix and ${\mathbf{T}}$ is the following transfer-matrix : \bea
{\mathbf{T}}=\left( \begin{array}{lcr}
      e^{\beta H_1}Z_+  & Z_0 \\
      Z_0  & e^{-\beta H_1}Z_- \label{T}
      \end{array}
\right). \eea The eigenvalues of ${\mathbf{T}}$ are \bea
\lambda_{\pm}=\frac{1}{2}\left(e^{\beta H_1}Z_+ +e^{-\beta H_1}Z_-
\pm \sqrt{\left(e^{\beta H_1}Z_+ -e^{-\beta H_1}Z_-
\right)^2+4Z_0^2} \right). \label{lamb}
 \eea
 Thus,
the free energy per one spin in the thermodynamic limit when only
maximal eigenvalue survives is \bea f=-\frac{1}{4 \beta}\log
\left(\frac{1}{2}\left(e^{\beta H_1}Z_+ +e^{-\beta H_1}Z_- +
\sqrt{\left(e^{\beta H_1}Z_+ -e^{-\beta H_1}Z_- \right)^2+4Z_0^2}
\right) \right), \label{fe_t}
 \eea
 For sublattice and total magnetization one obtains
 \bea
&& M_1=\frac{e^{\beta H_1}Z_+ -e^{-\beta
H_1}Z_-}{\sqrt{\left(e^{\beta H_1}Z_+ -e^{-\beta H_1}Z_-
\right)^2+4Z_0^2}}, \label{MMM} \\ \nonumber
 &&M_2=\frac{e^{\beta H_1}X_+ +e^{-\beta H_1}X_-+\frac{Z_+ \left(e^{\beta
H_1}X_+-X_-\right)+Z_-\left(e^{-\beta H_1}X_--X_+
\right)+4Z_0X_0}{\sqrt{\left(e^{\beta H_1}Z_+ -e^{-\beta H_1}Z_-
\right)^2+4Z_0^2}}}{3\left(e^{\beta H_1}Z_+ +e^{-\beta H_1}Z_- +
\sqrt{\left(e^{\beta H_1}Z_+ -e^{-\beta H_1}Z_- \right)^2+4Z_0^2}
\right)}, \\ \nonumber &&M=\frac{\langle\sum_{i=1}^N \left(\sigma_i+
\sum_{a=1}^3S_{ia}^z\right)\rangle}{4N}=\frac{1}{4}M_1+\frac{3}{4}
M_2,
 \eea
 where the functions $X_+,X_-,X_0$ are defined in the Appendix.
 If one assume $H_1=H_2=H$ which means the equality of the $g$-factors
 for $S$- and $\sigma$-spins the expression for the total magnetization
 per spin takes the following form:
 \bea
M=\frac{e^{\beta H}(Z_++X_+) -e^{-\beta H}(Z_--X_-)+\frac{(e^{\beta
H}Z_+ -e^{-\beta H}Z_-)(e^{\beta H}(Z_++X_+) +e^{-\beta
H}(Z_--X_-))+4Z_0X_0}{\sqrt{\left(e^{\beta H}Z_+ -e^{-\beta H}Z_-
\right)^2+4Z_0^2}}}{4\left(e^{\beta H}Z_+ +e^{-\beta H}Z_- +
\sqrt{\left(e^{\beta H}Z_+ -e^{-\beta H}Z_- \right)^2+4Z_0^2}
\right)}, \label{Mtot}
 \eea
 where $H_2=H$ should be put in all $X_{\alpha}$ and $Z_{\alpha}$
 functions.
 \section{Thermodynamics and magnetic properties.}
 In this section we examine the thermodynamic and magnetic
 properties of the model under consideration. Having the exact explicit expressions for the magnetization, specific heat, entropy and
 nearest neighbor correlation functions one can easily obtain their
 plot vs. temperature or external magnetic field, despite of their
 complicated  form. Also, analyzing the low--temperature region one
 can obtain the ground state properties and corresponding phase
 diagrams. In that follows, we restrict ourselves with the case of all
 antiferromagnetic couplings, $J>0$, $K>0$ and $\Delta>0$.
 \subsection{Zero temperature ground state phase diagrams.}
\begin{figure}
 \begin{center}
  \includegraphics[scale=0.5]{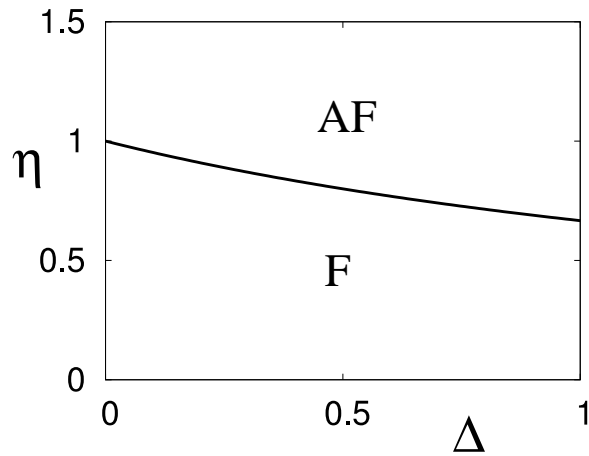}
  \caption{$T=0$, $H=0$ ground-state phase diagram in the $\Delta-\eta$ plane. The boundary between ferrimagnetic(F) and
  antiferromagnetic(AF)
  ground states are given by $\eta=\frac{2}{2+\Delta}$\label{fig2}}
           \end{center}
         \end{figure}

\begin{figure}
 \begin{center}
  \includegraphics[scale=0.4]{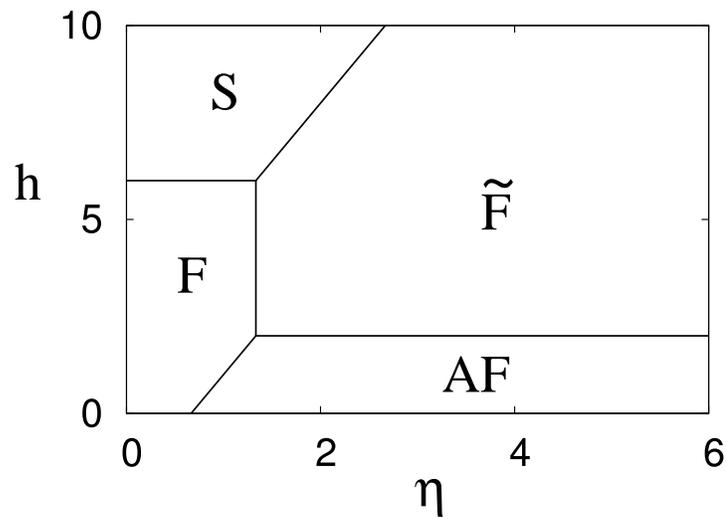}
  \caption{$T=0$ ground-state phase diagram in the $\eta-h$ plane for $\Delta=1$. The boundary between
  ferrimagnetic(F)
   and antiferromagnetic(AF) phases is $h=\eta(2+\Delta)-2$; between second ferrimagnetic phase($\widetilde{F}$) and saturates state(S)
   is $h=\eta(2+\Delta)+2$.\label{fig3}}
           \end{center}
         \end{figure}

 Let us analyze the possible ground state of the system at $T=0$
 and for positive values of all model parameters $J, K, \Delta$.
 According to the possible states of three $s=1/2$ spins one can
 distinguish the following spin configurations on the chain under
 consideration. First of all, the ground states in the absence of
  magnetic field depending on the values of model parameters can be
  either ferrimagnetic (F) or antiferromagnetic (AF),
  \bea
 && |F
  \rangle=\prod_{i=1}^N|3/2,3/2\rangle_i\bigotimes|\downarrow\rangle_i,
  \label{states1} \\ \nonumber
   &&|AF\rangle=\prod_{i=1}^N|1/2,1/2\rangle_i\bigotimes|\downarrow\rangle_i,
  \eea
where $|l,m\rangle_i$ stands for the spin state of triangle at
$i$-th plaquette with total spin equal to $l$ and $S^z=m$ and
$|\uparrow\rangle, |\downarrow\rangle$ are the standard "up" and
"down" states of the Ising spins. Expanding the states of quantum
triangles by the single spin basis one finds \bea
&&|3/2,3/2\rangle=|\uparrow\uparrow\uparrow\rangle, \label{states2}
\\ \nonumber
&&|1/2,1/2\rangle_{R,L}=\frac{1}{\sqrt{3}}\left(|\uparrow\uparrow\downarrow\rangle
+e^{\pm \frac{2 \pi i}{3}}|\uparrow\downarrow\uparrow\rangle +e^{\mp
\frac{2 \pi i}{3}}|\downarrow\uparrow\uparrow \rangle \right), \label{hvs}
 \eea
 where the choice of the sings of the coefficients in
 $|1/2,1/2\rangle$ state is connected with the its chirality. For
 the interactions considered here these two states are degenerated
 by energy
 and where is no need to distinguish them in our consideration, so
 we will omit $R,L$ indices. Thus, here the manifestation of frustration
consists in this degeneracy. Every triangle can spontaneously pass
from $|1/2,1/2\rangle_R$ to $|1/2,1/2\rangle_L$ and vice versa.
Therefore, due to the frustration the system possesses non zero
entropy at $T=0$. It is worth mentioning, that considering the Ising
limit of the system, $\Delta=0$, one can obtain essential
enhancement of frustration at $J=K$. Indeed, if one takes only one
plaquette which consists of five Ising spins with nine uniform couplings
between them, the ground state will be disordered one with three
spins pointing up and the remaining two pointing down or vise versa. So,
the number of possible degenerated configuration for one plaquette
and for fixed orientation of the total magnetization will be 10. The
 energies per one plaquette and magnetizations corresponding to $|F\rangle$ and $|AF\rangle $ configurations are:
\bea &&\varepsilon_F=3\eta-2 h-6, \quad M_F=1/2 \label{en1} \\
\nonumber &&\varepsilon_{AF}=-\eta(1+2\Delta)-2, \quad M_{AF}=0
 \eea
 where the following dimensionless parameters are introduced, $\varepsilon=E/N
 K$, $\eta=J/K$, $h=H/K$. Thus, the ground state of the system at
 $T=0$ and $h=0$ is ferrimagnetic when $\eta \leq
 \frac{2}{2+\Delta}$ and antiferromagnetic otherwise. The
 corresponding phase diagram is presented in Fig. (\ref{fig2}).
 The appearance of the external magnetic field makes another two ground
 states possible. Namely,  the states with total magnetization $M=1/2$
 resulting from AF-state by the flip of all Ising spins, which actually is a ferrimagnetic
 state differing from the (F) and
 completely polarized saturated phase:
  \bea
    &&|\widetilde{F}\rangle=\prod_{i=1}^N|1/2,1/2\rangle_i\bigotimes|\uparrow\rangle_i,   \label{states3} \\ \nonumber
 && |S
  \rangle=\prod_{i=1}^N|3/2,3/2\rangle_i\bigotimes|\uparrow\rangle_i,
  \eea
   with
 \bea
&& \varepsilon_{\widetilde{F}}=-\eta (1+2 \Delta)-2 h+2, \quad
M_{\widetilde{F}}=1/2,
 \label{en2} \\ \nonumber
 &&\varepsilon_{S}=3 \eta-4 h+6, \quad M_S=1.
 \eea
 The corresponding $T=0$ ground state phase diagram in the $(h,\eta)$
 plane for the case $\Delta=1$ is presented in Fig. (\ref{fig3}).
 \subsection{Magnetization processes and susceptibility}
\begin{figure}
 \begin{center}
  \includegraphics[scale=0.7]{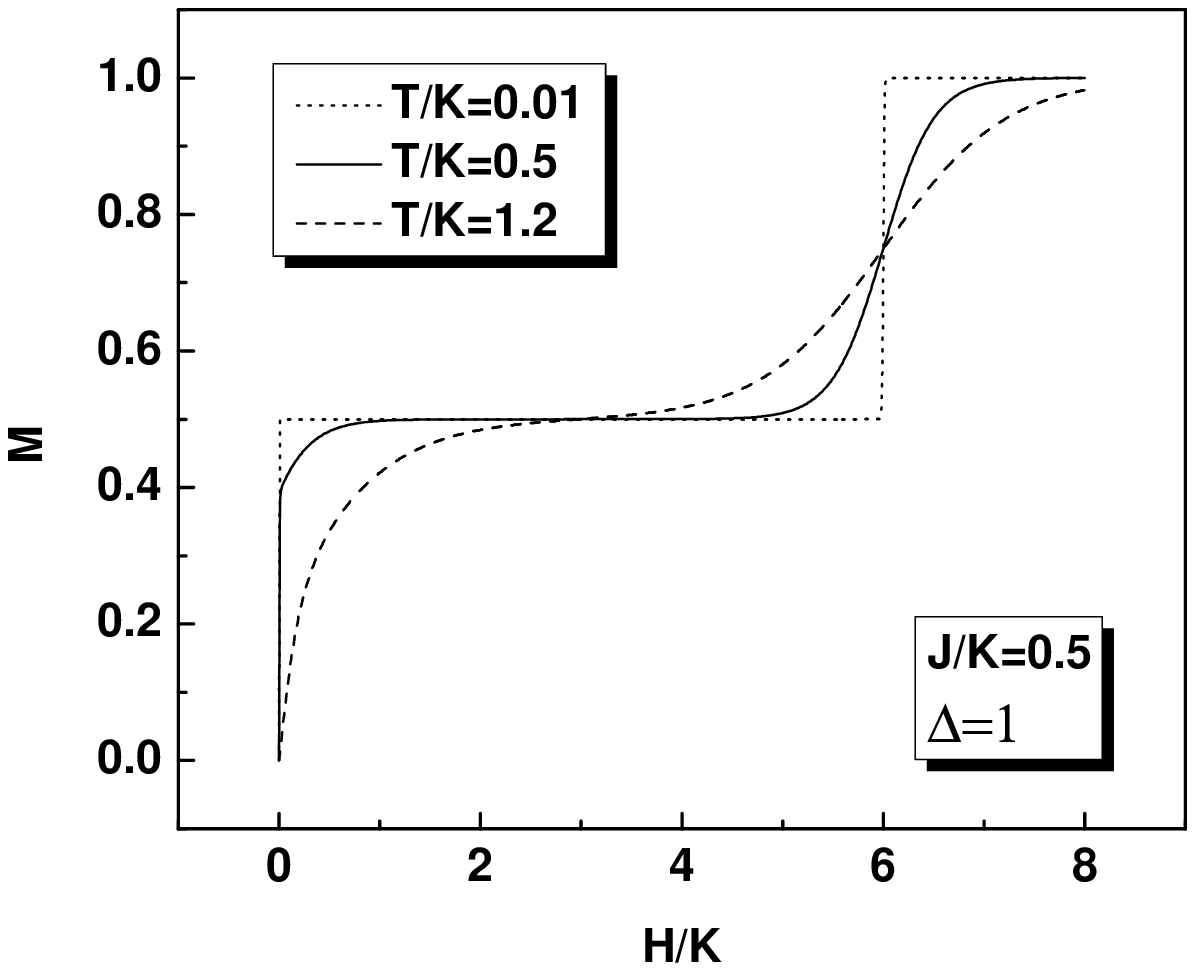}
  \caption{Total magnetization $M$ as a function of dimensionless external magnetic field $h$ for $\eta=0.5, \Delta=1$ and three different dimensionless temperatures
  $T/K=0.01,0.5$ and $1.2$.\label{fig4}}
           \end{center}
         \end{figure}

\begin{figure}
 \begin{center}
  \includegraphics[scale=0.7]{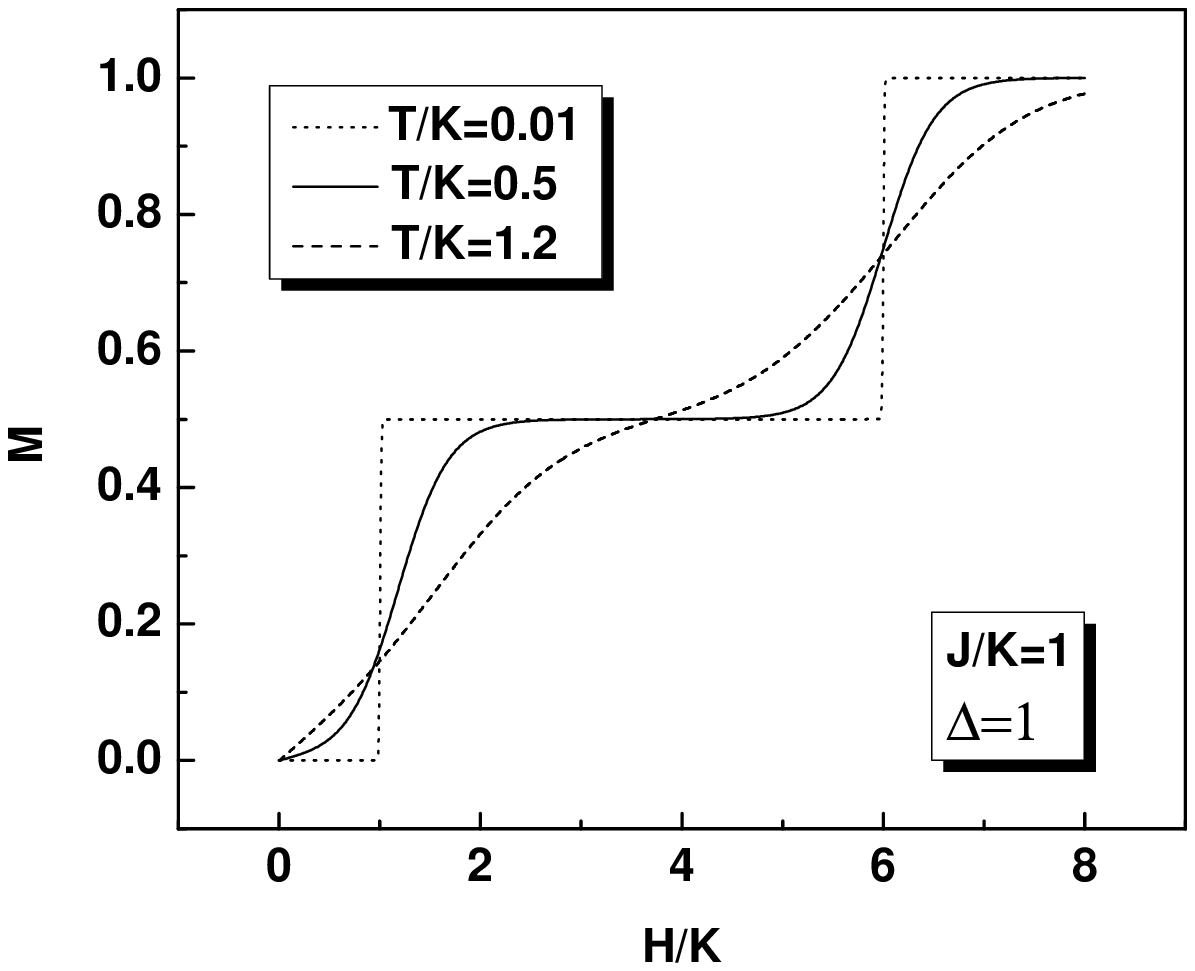}
  \caption{Total magnetization $M$ as a function of dimensionless external magnetic field $h$ for $\eta=1, \Delta=1$ and three different dimensionless temperatures
  $T/K=0.01,0.5$ and $1.2$..\label{fig5}}
           \end{center}
         \end{figure}

\begin{figure}
 \begin{center}
  \includegraphics[scale=0.7]{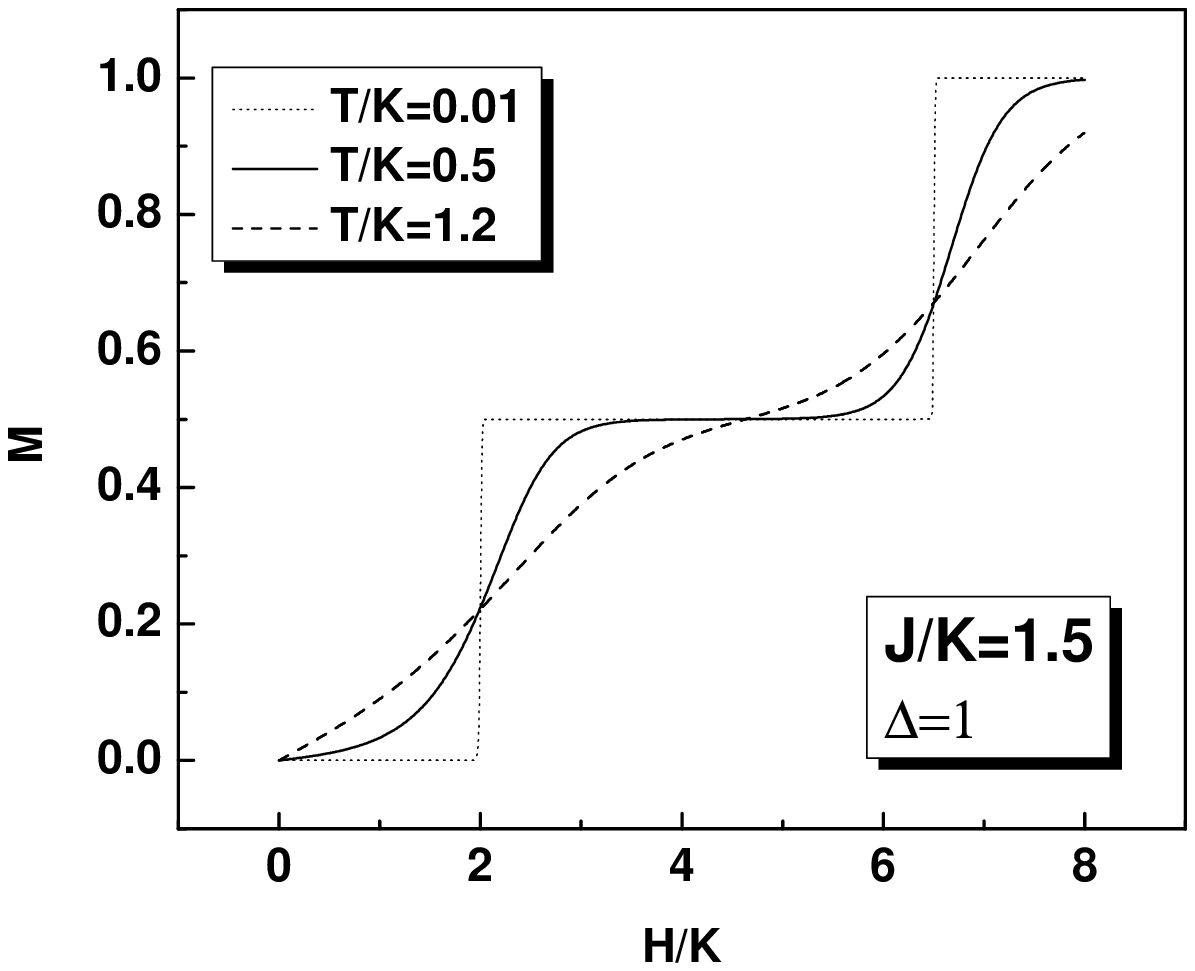}
  \caption{Total magnetization $M$ as a function of dimensionless external magnetic field $h$ for $\eta=1.5, \Delta=1$ and three different dimensionless temperatures
  $T/K=0.01,0.5$ and $1.2$.\label{fig6}}
           \end{center}
         \end{figure}

 According to the phase diagram displayed in Fig. (\ref{fig3}) one can expect three different kinds of magnetic
 behavior depending on the relation between parameters $\eta$ and
 $\Delta$. The following sequences of transitions occur when the magnitude of external magnetic
 field is increasing. For the values of $\eta$ and $\Delta$ belonging
 to the first region given by $
 \eta \leq \frac{2}{2+\Delta}$, the system undergoes only one transitions from $|F\rangle$ to $|S\rangle$
 taking place at $h=6$. The next range of parameters values is characterized
 by $\frac{2}{2+\Delta}< \eta \leq \frac{4}{2+\Delta}$. Here one can
 occur two consecutive
 transitions from $|AF\rangle$ to $|F\rangle$ at $h=\eta
 (2+\Delta)-2$ and from $|F\rangle$ to $|S\rangle$ at $h=6$. And,
 finally, when $\eta$ is greater than $\frac{4}{2+\Delta}$ the
 system undergoes the following transitions $|AF \rangle \rightarrow |\widetilde{F}\rangle$ and $|\widetilde{F}\rangle \rightarrow |S\rangle $
at $h=2$ and $h=\eta(2+\Delta)+2$ respectively. The corresponding
plots of the magnetization processes for finite temperatures are
shown in Fig.(\ref{fig4})-(\ref{fig6}). At zero temperature all
transitions are obviously jump-like and the corresponding
magnetization curves are perfectly step-like. However, at arbitrary
finite temperatures all transitions are smeared out within some
interval of magnetic field. In Fig.(\ref{fig4}) the plots of $M$ vs
$H/K$ are presented for the $\eta=0.5$ and $\Delta=1$ for several
temperatures. One can see the magnetization plateau at $M=1/2$ which
corresponds to the zero temperature stability of $|F \rangle$ state
in the interval of $h$ from $0$ to $6$. The plot for $T/K=0.01$ is
very close to this picture, whereas one can see essential shrinkage of
the plateau width with increasing temperature. Surely, for
large enough temperatures the magnetization curve acquires the form
corresponding to the free spins (Langevin curve) with monotonic
increase between 0 and 1 values. Generally speaking, at the small
values of the ration $\eta$ the system under consideration is
similar to the
 mixed antiferromagnetic Ising chain with alternating $s=1/2$ and $s=3/2$
spins, because the strong interaction between the triangle spins
makes their behavior very similar to that of one spin-$3/2$. So, the
region $\eta \leq \frac{2}{2+\Delta}$ can be considered as the
spin-3/2-spin-1/2 mixed chain regime. The typical plots for the
second range of the $\eta$ is presented in Fig. (\ref{fig5}). Here
two magnetization jumps  corresponding to the transitions
$|AF\rangle \rightarrow |F \rangle$ and $|F \rangle \rightarrow |S
\rangle$ occur at $T=0$ in the magnetization curves for the system
under consideration for $\eta=1$ and $\Delta=1$. The corresponding
plateau at $M=1/2$ as in the previous case appears due to stability
of the $|F \rangle$ state between $|AF\rangle$ and $|S \rangle$. The
width of the plateau is $8-\eta (2+\Delta)$. The plots for different
finite temperatures from Fig. \ref{fig5} exhibit the gradual
shrinking down of the plateau with increasing temperature. The
same plateau at $M=1/2$ but with another physical content one can
see in the magnetization curves for the third region of values of
$\eta$. In the Fig. (\ref{fig6}) one can see the plots of $M$ vs.
$H/K$ for $\eta=1.5$ and $\Delta=1$. All properties mentioned above
concerning the temperature effect holds in this case as well.
However, the plateau at $M=1/2$ now corresponds to the
$|\widetilde{F} \rangle$ state of the chain. The two terminal points
of the plateau correspond to the zero temperature transition from
$|AF \rangle$ to $|\widetilde{F} \rangle$ and from  $|\widetilde{F}
\rangle$ to $|S\rangle$ respectively. The position of the left end
of the plateau ($h=2$) is unaffected by the any changes of
interaction parameters provided the later stay within the $\eta >
\frac{4}{2+\Delta}$.( The same phenomenon for the right end of
plateau at $h=6$ take place in the two other regimes of behavior of
the system under consideration.) The width of this plateau at $T=0$
equal to $\eta (2+\Delta)$.

\begin{figure}
 \begin{center}
  \includegraphics[scale=0.7]{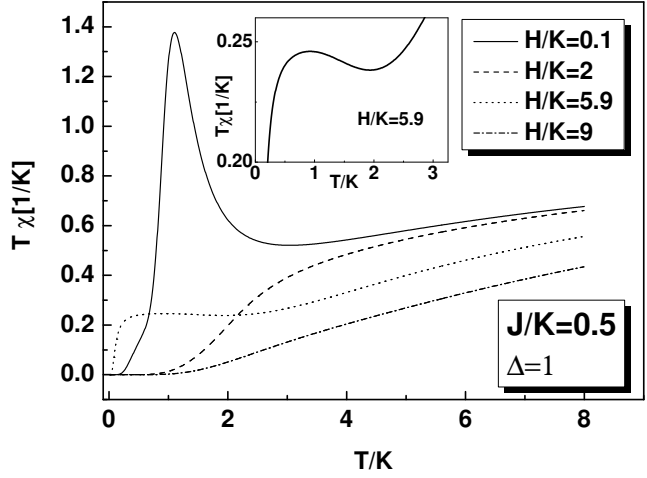}
  \caption{The temperature dependence of the $T \chi$ product for several values of field strength and $\eta=0.5, \Delta=1$ Inset displays the round minimum for $H/K=3$ on an enlarge scale .\label{fig7}}
           \end{center}
         \end{figure}

\begin{figure}
 \begin{center}
  \includegraphics[scale=0.7]{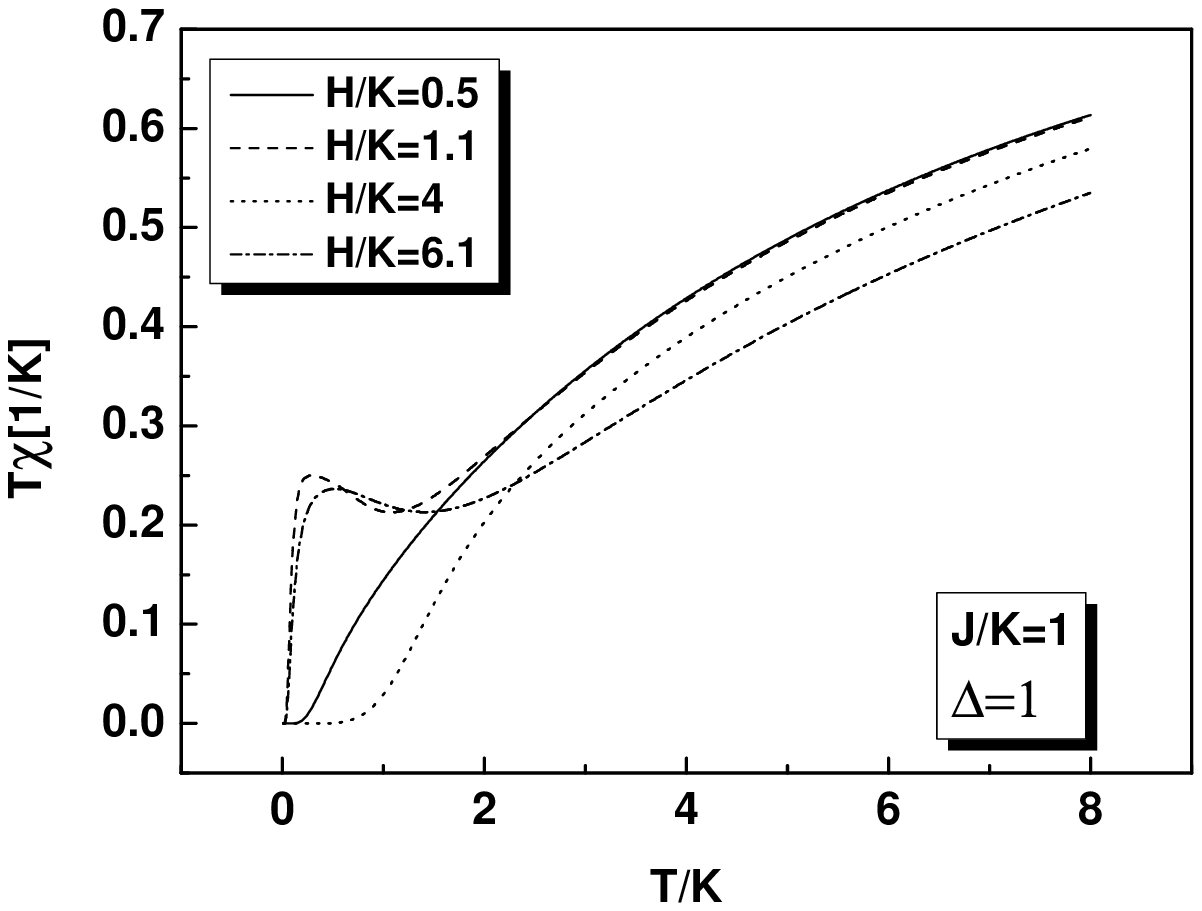}
  \caption{The temperature dependence of the $T \chi$ product for several values of field strength and $\eta=1, \Delta=1$ .\label{fig8}}
           \end{center}
         \end{figure}

\begin{figure}
 \begin{center}
  \includegraphics[scale=0.7]{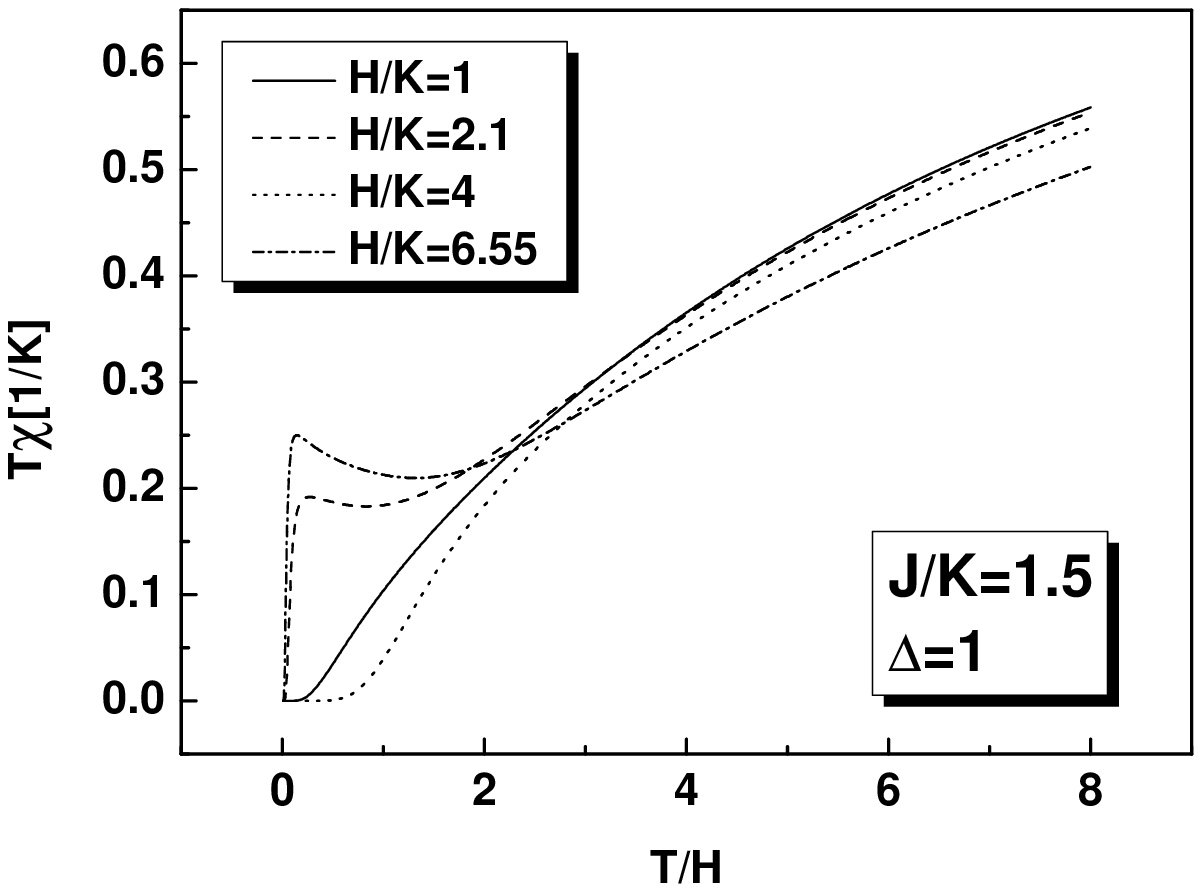}
  \caption{The temperature dependence of the $T \chi$ product for several values of field strength and $\eta=1.5, \Delta=1$ .\label{fig9}}
           \end{center}
         \end{figure}

    The plots of thermal dependence of magnetic susceptibility times
    temperature $T\chi$ per one spin are displayed in the Figs.
    (\ref{fig7})-(\ref{fig9}). First of all, zero-field
    susceptibility diverges at $T=0$ which is a consequence of the
    fact that $T=0$ is the critical point at which the system
    possess an ideal long range spin order, which is destroyed by any finite
    temperature. In Fig. (\ref{fig7}) the plots of $T\chi$ vs. $T$   for
    the parameters first region are presented for several values of
    external magnetic field. Here no divergence is observed at $T=0$
    because of energy gap opened by magnetic field. A rather sharp
    peak appears in the curve at extremely weak field value. This
    peak corresponds to the thermal instability at the transition
    from uncorrelated disordered state, which is the ground state at any finite temperature
    and $H=0$, to the ferrimagnetic state $|F \rangle$.
    At the end of the peak the tiny round minimum is
    observed. With increasing of the field strength the low
    temperature peak is smoothed out and the curve become
    monotonically increasing, which is general feature of
    antiferromagnetically coupled spin systems. At $H/K=5.9$,
    i.e. in the vicinity of the transition from  $|F\rangle$ to $|S\rangle$ the $T\chi$ vs. $T$
    curve exhibits a very rapid increase at low temperatures with
    further narrow plateau formation which is then followed by the
    slowly and monotonically increasing region. The inset of Fig. (\ref{fig7}) shows
    the small round minimum which accompanies the transition from
    narrow plateau to increasing monotonic  behavior of
    susceptibility. Very similar picture of susceptibility thermal
    dependence was obtained for the F-F-AF-AF Ising-Heisenberg chain
    in Ref. \cite{str2}. In Fig. (\ref{fig8}) the plots of $T\chi$ vs.
    $T$ for $\eta=1, \Delta=1$ are presented. Here, a little smaller
    than in the previous case, low temperature peaks appear at the
    values of $H/K$ corresponding to the transitions from $|AF
    \rangle$ to $|F \rangle$ and from $|F \rangle$ to $|S \rangle$.
     Similar picture one can observe for the $T\chi$
    thermal behavior for the system under consideration when the
    parameters belong to the third region, $\eta >
    \frac{4}{2+\Delta}$. The corresponding plots are presented in
    Fig. (\ref{fig9}).
    \subsection{Specific heat}
\begin{figure}
 \begin{center}
  \includegraphics[scale=0.7]{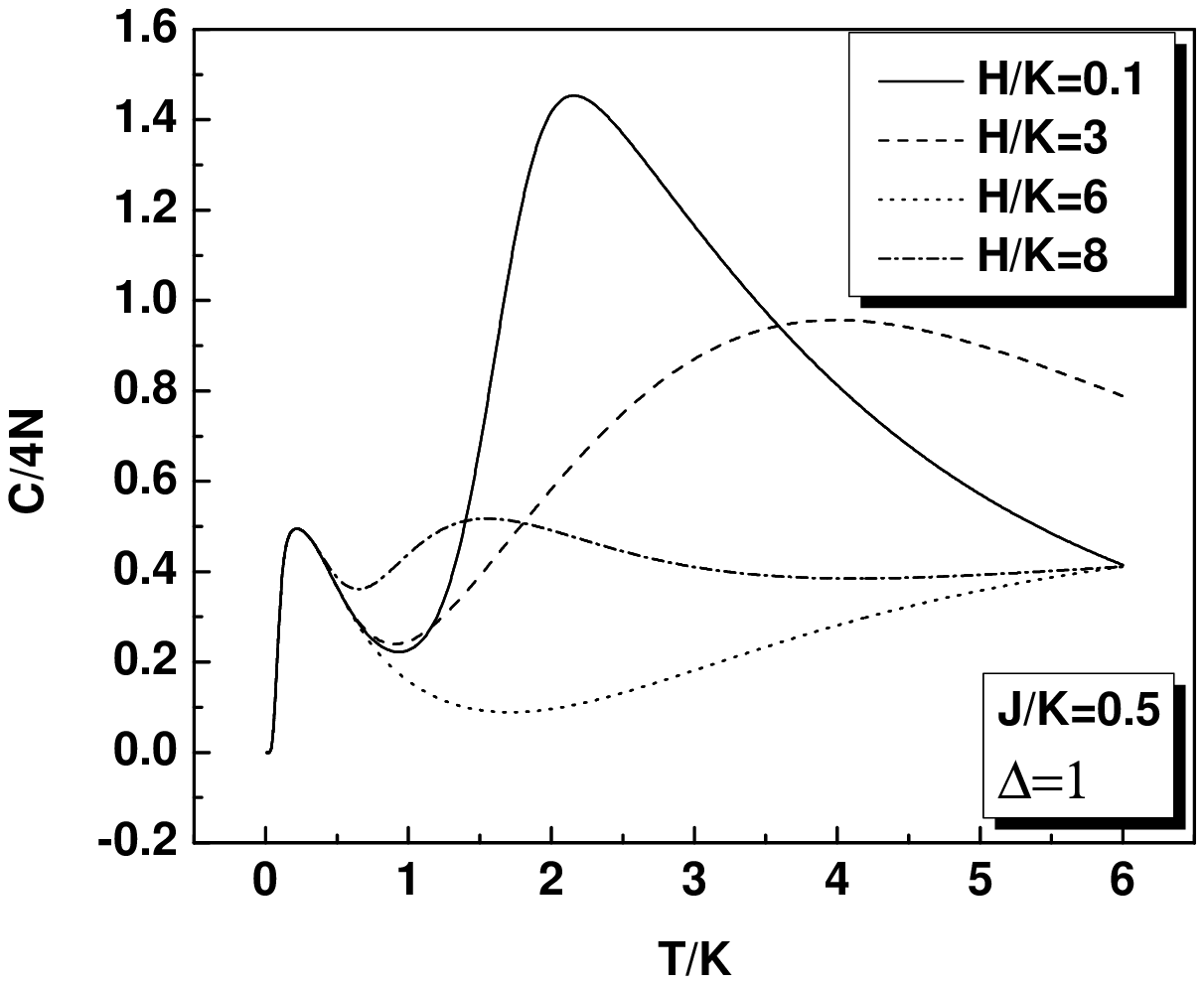}
  \caption{Specific head per one spin as a function of the dimensionless temperature $T/K$ given in $K$ units for
  $\eta=0.5, \Delta=1$ for various values of $h$. \label{fig10}}
           \end{center}
         \end{figure}

\begin{figure}
 \begin{center}
  \includegraphics[scale=0.7]{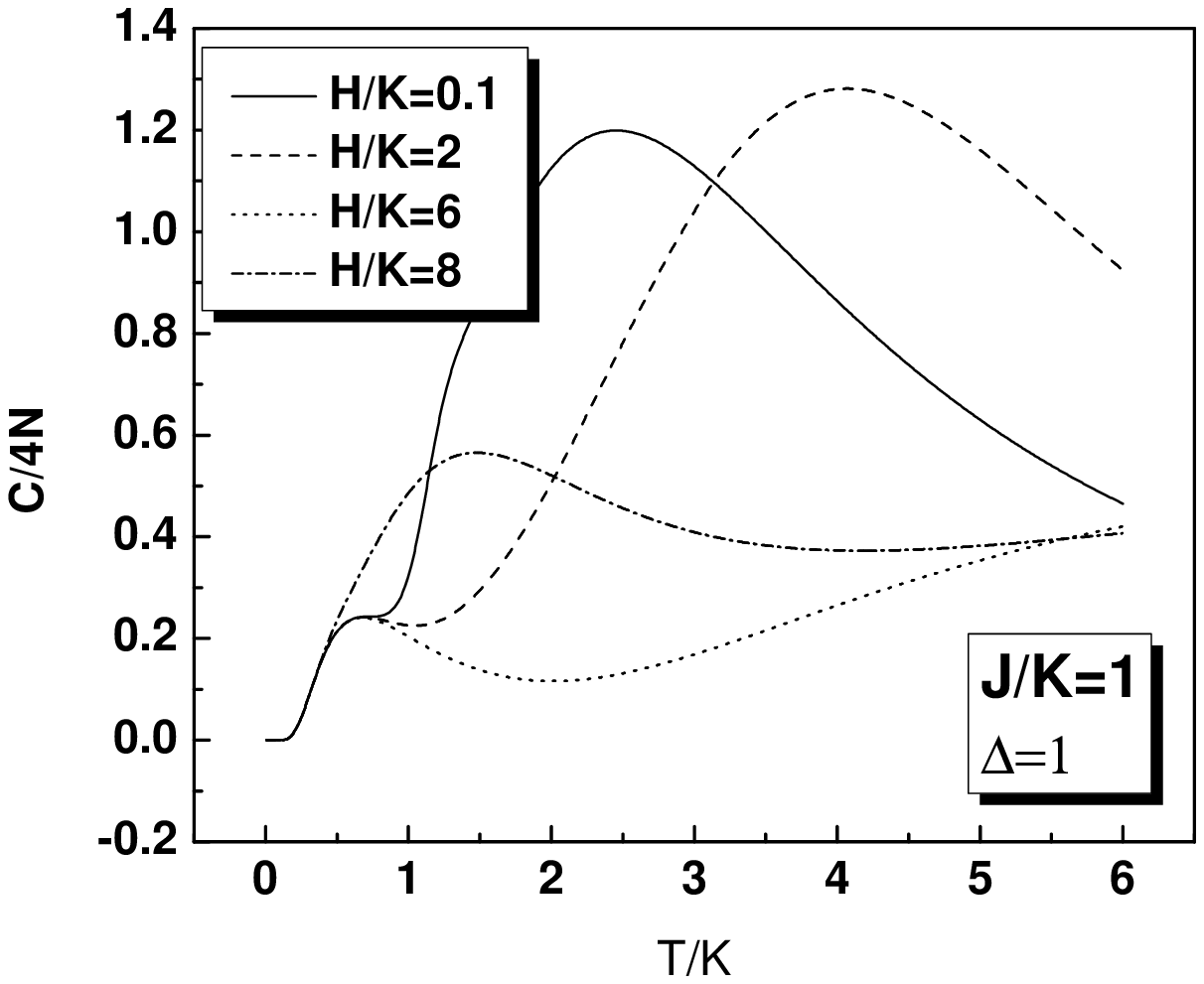}
  \caption{Specific head per one spin as a function of the dimensionless temperature $T/K$ given in $K$ units for
  $\eta=1, \Delta=1$ for various values of $h$.\label{fig11}}
           \end{center}
         \end{figure}

\begin{figure}
 \begin{center}
  \includegraphics[scale=0.7]{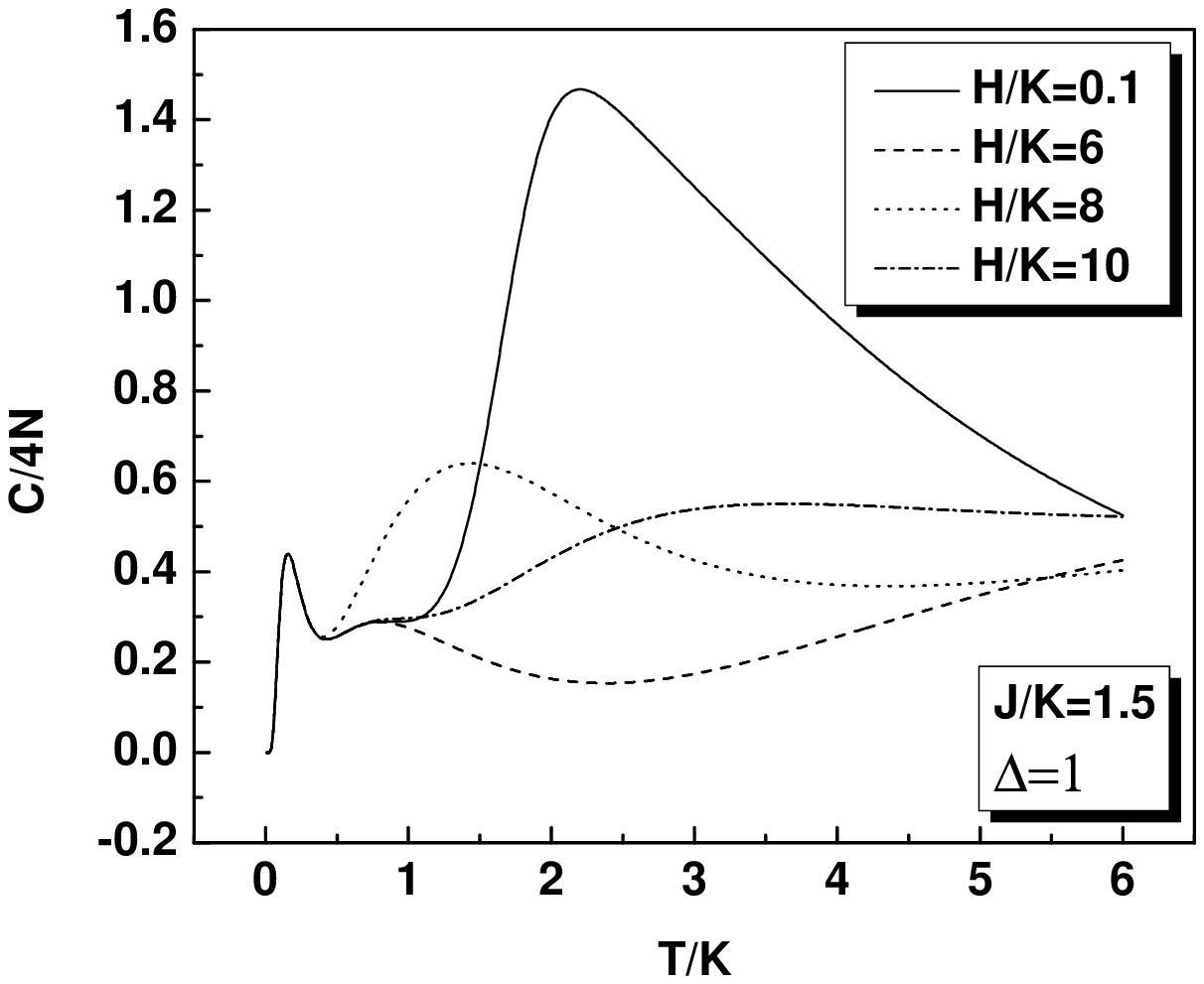}
  \caption{Specific head per one spin as a function of the dimensionless temperature $T/K$ given in $K$ units for
  $\eta=1.5, \Delta=1$ for various values of $h$.\label{fig12}}
           \end{center}
         \end{figure}

    As mentioned above, the technique developed in the paper allows
    one to obtain analytical exact expressions for all thermodynamic
    functions of the system. Thought, the forms of some  of them are
    extremely cumbersome, one can easily obtain their plots. In
    Figs. (\ref{fig10})-(\ref{fig12}) the plots of thermal behavior
    of the specific heat of the system under consideration are
    presented. Generally, specific heat exhibits two-peak structure,
with one sharp low temperature, which is almost unaffected
by the value of the magnetic field, and another one
with broaden peak, which undergoes considerable changes
under increasing magnetic field. Fig. (\ref{fig10}) demonstrates the
    corresponding plots for $\eta=0.5, \Delta=1$. The second peak is
    well pronounced here and is strongly field dependant. Thus, one can associate it with Schottky-type anomaly. With the increase in the external magnetic
    field strength the peak moves to higher temperature region
    and, at the same time,  drops in magnitude. At the values of
    field strength close to the transition to saturated state at
    $T=0$ the second broad peak begins to enhance again for a while,
    however  with the further increase in the field it gradually
    gets smoother. In general, one can
    observe almost the same features  in the thermal behavior of the specific heat times temperature per spin
    for the region $ \frac{2}{2+\Delta}\leq \eta
    <\frac{4}{2+\Delta}$ (Fig.\ref{fig11}). The essential difference
    with the previous case is completely merging of sharp
    low-temperature peak and the Schottky-type broad peak at the
    values of field strength corresponding to the plateau in the
    magnetization curve at $T=0$. Fig. (\ref{fig12}) shows typical plots of $T\chi$ vs.
    $T$ for $\eta=1.5$, i. e. for the third region of model
    parameters, $\eta \geq \frac{4}{2+\Delta}$. Here at low field
    strength immediate after the first sharp peak a narrow almost
    horizontal region of the curve appears. Increasing the
    field strength one can observe the enhancement of the slope of the
    corresponding region accompanying the second broad peak
    changes described above.
    \section{Concluding remarks}

In this paper we considered exactly solvable model of one
dimensional spin system with Heisenberg $XXZ$ triangular spin
clusters alternating with single Ising spins. Two different
approaches have been discussed, the decoration--iteration
transformation \cite{str2, str, mt} and the transfer-matrix direct
formalism. Both approaches lead to the possibility of obtaining
exact analytical expressions for the thermodynamic functions of the
system. The system considered here, as well as the other systems
considered earlier in Ref. \cite{ffa}-\cite{Iran} at first glance
is only of academic interest. However, the strict indications of
the relevance of the approaches based on the simplification of the
spin exchange interaction scheme in one-dimensional magnetic
materials have been given in Ref. \cite{ffa, str2, ayd}. Namely,
considering the spin exchange Hamiltonians for various materials
with one--dimensional exchange structure in most cases one can
gain insight in description of their properties and understanding
of their thermodynamic behavior only within the complicated and
laborious numerical calculations demanding some time extremely
powerful computing facilities, such as supercomputers. On the
other hand, at least for some class of systems, the simplification
of the model based on the replacement of all or just some of the
exchange Heisenberg interactions with the interactions of Ising
type, does not affect the prominent properties of their
thermodynamics. This fact makes the simplified exactly solvable
(in the "thermodynamic" sense) counterpart of the one-dimensional
models which describes the magnetism of real materials very
promising. Especially this approach can be successful in the
series of bond alternating chain with spatially repeated sequence
of ferromagnetic and antiferromagnetic interactions as it has been
established for F-F-AF \cite{ffa} and F-F-AF-AF\cite{str2} chains.
Apparently, replacing the ferromagnetic bounds by Ising ones is
especially "safe" in that sense, because the former ones favor
parallel orientation of the spins, which is given by a separable
vector of state. Thus, one can lose only a small amount of
information of the properties of that bound by regarding it as the
Ising one. Antiferromagnetically coupled spins are not so
"harmless", because they can be in the superposition and entangled
states which cannot be properly translated into the Ising
language. Another essential difference between Heisenberg spin
chains and their simplifying counterparts is the spin-wave
properties. Generally speaking, spin wave cannot propagate through
 Ising spins, so only localized excitations are relevant in that
case. Further applications of the methods discussed in the present
paper in other one--dimensional and quasi--one--dimensional
Heisenberg models, especially in the models of novel magnetic
materials, as well as a comparison of the results with the
experimental data, can play a very important role in the
understanding of complicated magnetic materials and their
thermodynamic properties and can draw the frameworks of
applicability of the approach offered in this paper.

\section{Acknowledgements}
We are indebted to J. Stre\v{c}ka and L. \v{C}anov\'{a} for bringing
to our attention the results of their researches on solving the
Ising--Heisenberg spin systems within decoration--iteration
transformation and for valuable discussions. We also express our
gratitude to T. Hakobyan for stimulating comments and interest
toward the paper and to O. Derzhko and L. Ananikian for fruitful
conversations. Special thanks are due to A. Valishev and A. Badasyan for help in preparing
the figures. V.O. express his gratitude for hospitality to
 LNF--INFN where the paper was finished.  This work was partly
supported by the European Community Human Potential Program under
contract MRTN-CT-2004-005104 \textit{``Constituents, fundamental
forces and symmetries of the universe''} and by 
CRDF-UCEP Grants No.07/06 and 07/02, ANSEF Grant No. 1386-PS and INTAS Grant No.05-7928.
 \section{Appendix}
 Here we derive the exact expressions for the functions
 $Z_+,Z_-,Z_0$ from Eq. (\ref{zzz}). For these purposes one should
 obtain the eigenvalues of the matrix, corresponding to the
 Hamiltonian of single triangle. So, we define a partition function
 for one single triangle
 \bea
{ \mathcal{Z}}_{\triangle}(\beta;J,\Delta, H)=\mbox{Tr} e^{-\beta
{\mathcal{H}}_{\triangle}(J, \Delta, H)}=\sum_{n=1}^8\langle
\psi_n|e^{-\beta {\mathcal{H}}_{\triangle}(J, \Delta,
H)}|\psi_n\rangle, \label{z_tr}
 \eea
where $|\psi_n\rangle$ is an arbitrary orthogonal normalized system of
states. Then, one can easily see that the Hamiltonian corresponding to
i-th plaquette from Eq. (\ref{H}) has the same form as the
Hamiltonian of single isolated triangle if one introduces instead of
magnetic field $H$ an "effective field" depending on the values of
$\sigma_i$ and $\sigma_{i+1}$: \bea
\tilde{H}=H_2-K(\sigma_i+\sigma_{i+1}). \label{eff}
 \eea
 The term corresponding to the interaction of $\sigma_i$ and
 $\sigma_{i+1}$ with magnetic field $H_1$ adds $-\frac{H_1}{2}(\sigma_i+\sigma_{i+1})$ to each eigenvalue of ${\mathcal{H}}_{\triangle}$.
 It is easy to see that ${\mathcal{H}}_{\triangle}$ is diagonal in the symmetry-adapted basis the higher weight state of which is given by Eq. (\ref{hvs}). The rest 5 states can be easily found by successive action of the lowering operator $S_{tot}^-=\frac{1}{2}(S_1^- +S_2^- +S_3^-)$ on $|3/2,3/2\rangle$ and $|1/2,1/2\rangle_{L,R}$:
 \bea
&& |3/2,1/2\rangle = \frac{1}{\sqrt{3}}\left(|\uparrow\uparrow\downarrow\rangle+|\uparrow\downarrow\uparrow\rangle+|\downarrow\uparrow\uparrow\rangle \right),\\ \nonumber
 &&|3/2,-1/2\rangle=\frac{1}{\sqrt{3}}\left(|\uparrow\downarrow\downarrow\rangle+|\downarrow\uparrow\downarrow\rangle+|\downarrow\downarrow\uparrow\rangle \right),\\ \nonumber
 && |3/2,-3/2\rangle=|\downarrow\downarrow\downarrow\rangle, \\ \nonumber
 && |1/2,-1/2\rangle_{R,L}=\frac{1}{\sqrt{3}}\left(|\downarrow\downarrow\uparrow\rangle +e^{\pm\frac{ i2 \pi}{3}}|\uparrow\downarrow\downarrow\rangle+e^{\mp \frac{i2 \pi}{3}}|\downarrow\uparrow\downarrow\rangle \right).
 \eea

 Taking into account Eq. (\ref{eff}) the eight eigenvalues
 of the ${\mathcal{H}}_i$ are
 \bea
&& \lambda_{1,2}(\sigma_i, \sigma_{i+1})=3(J\pm H)\mp
 3K(\sigma_i+\sigma_{i+1}), \label{eig} \\ \nonumber
 &&\lambda_{3,4}(\sigma_i,
 \sigma_{i+1})=\lambda_{5,6}(\sigma_i, \sigma_{i+1})=-(1+2\Delta)J \pm H \mp
 K(\sigma_i+\sigma_{i+1}), \\ \nonumber
 &&\lambda_{7,8}(\sigma_i, \sigma_{i+1})=-(1-4\Delta)J \pm H \mp
 K(\sigma_i+\sigma_{i+1}).
 \eea
 Thus, one have
 \bea
Z(\sigma_i, \sigma_{i+1})=\sum_{i=1}^8 e^{-\beta \lambda_n(\sigma_i,
\sigma_{i+1})}=B_1
\cosh(\beta(H_2-K(\sigma_i+\sigma_{i+1})))+B_2\cosh(\beta3(H_2-K(\sigma_i+\sigma_{i+1}))),
\label{zss}
 \eea
 where
 \bea
 B_1=2\left(e^{\beta(1-4 \Delta) J}+2e^{\beta(1+2 \Delta) J}
 \right), \quad B_2=2e^{-\beta 3 J}. \label{BB}
 \eea
 Now, we can define the functions appearing in Eqs.
 (\ref{av})-(\ref{cor}):
 \bea
 &&Z_+=B_1\cosh(\beta(H_2-2K))+B_2\cosh(\beta3(H_2-2K)),
 \label{func} \\ \nonumber
 &&Z_-=B_1\cosh(\beta(H_2+2K))+B_2\cosh(\beta3(H_2+2K)), \\
 \nonumber
 &&Z_0=B_1\cosh(\beta H_2)+B_2\cosh(\beta3H_2), \\ \nonumber
 &&X_{\alpha}=T\left(\frac{\partial Z_{\alpha}}{\partial H_2}
 \right)_T= B_1 \sinh(\beta  (H_2-\alpha 2 K))+3B_2 \sinh(\beta 3(H_2-\alpha 2K)), \quad \alpha=\pm, 0, \\ \nonumber
 &&Y_{\pm}=T\left(\frac{\partial Z_{\pm}}{\partial K} \right)_T=\mp 2 \left(B_1 \sinh (\beta (H_2 \mp 2K))+3 B_2 \sinh (\beta 3(H_2 \mp 2K)) \right), \\
 \nonumber
 &&Q_{\alpha}=T \left(\frac{\partial Z_{\alpha}}{\partial \Delta}
 \right)_T=B_3 \cosh(\beta(H_2 - \alpha 2 K)), \quad \alpha= \pm,0, \\ \nonumber
 && P_{\alpha}=T \left(\frac{\partial Z_{\alpha}}{\partial J}
 \right)_T=B_4\cosh(\beta  (H_2-\alpha 2 K)) + B_5\cosh(\beta 3(H_2-\alpha 2K)), \quad \alpha= \pm,0, \\ \nonumber
 &&W_{\pm}=\frac{X_+}{Z_+}+\frac{X_-}{Z_-}\pm 2\frac{X_0}{Z_0},
 \quad W_0=\frac{X_+}{Z_+}-\frac{X_-}{Z_-}, \\ \nonumber
 &&V_{\pm}=\frac{Y_+}{Z_+}\pm \frac{Y_-}{Z_-}, \\ \nonumber
 &&U_{\pm}=\frac{Q_+}{Z_+}+\frac{Q_-}{Z_-}\pm 2\frac{Q_0}{Z_0},
 \quad U_0=\frac{Q_+}{Z_+}-\frac{Q_-}{Z_-}, \\ \nonumber
 && F_{\pm}=\frac{P_+}{P_+}+\frac{P_-}{Z_-}\pm 2\frac{P_0}{Z_0},
 \quad F_0=\frac{P_+}{Z_+}-\frac{P_-}{Z_-},
 \eea
 where
 \bea
&&B_3=16 J  e ^{\beta(1-\Delta) J} \sinh (\beta 3 \Delta J) \label{BBB}\\
\nonumber
 &&B_4=2\left((1-4 \Delta) e ^{\beta(1-4 \Delta) J}+2(1+2\Delta) e ^{\beta(1+2 \Delta) J}
 \right), \\ \nonumber
 &&B_5=-6 e^{-\beta 3 J}.
 \eea

\newpage


\end{document}